\documentclass[aps,onecolumn,amsmath,amssymb]{revtex4}
\usepackage{color}
\usepackage{graphics}
\usepackage{epsfig}
\usepackage{amsmath}
\usepackage{amssymb}
\usepackage{amsthm}
\usepackage[mathscr]{eucal}
\usepackage{graphicx}
\usepackage{dcolumn}
\usepackage{bm}

\newcommand{\Eq}[1]{{Eq.~({\ref{#1}})}}

\newcommand{\bea}{\begin{eqnarray}}
\newcommand{\eea}{\end{eqnarray}}
\newcommand{\beas}{\begin{eqnarray*}}
\newcommand{\eeas}{\end{eqnarray*}}
\newcommand{\sumint}{\sum\!\!\!\!\!\!\!\!\int}

\begin{document}
\title{Quarkyonic Chiral Spirals in a Magnetic Field}

\author{Efrain J. Ferrer, \underline{Vivian de la Incera}\footnote{Talk presented at HIC for FAIR Workshop and XXVIII Max Born Symposium "Three Days on Quarkyonic Island", Wroclaw, Poland, May 19-21, 2011} and Angel Sanchez}
\affiliation{ Department of Physics, University of Texas at El Paso,
  500 W. University Ave., El Paso, TX 79968, USA}
\date{\today}

\begin{abstract}
We discuss the formation of quarkyonic chiral spirals in the presence of a magnetic field. The explicit breaking of the rotational symmetry by the external magnetic field gives rise to an additional chiral spiral that varies along the field direction and rotates in the chiral space between pion and magnetic moment components.
\end{abstract}

\maketitle

\section{Introduction}

Quarkyonic matter (QyM) is a large $N_c$ phase of cold dense quark matter recently suggested by McLerran and Pisarski \cite{quarkyonic-matter}. The main feature of QyM is the existence of asymptotically free quarks deep in the Fermi sea and confined excitations at the Fermi surface. The quarks lying deep in the Fermi sea are weakly interacting because they are hard to be excited due to Pauli blocking. Their interactions are hence very energetic and the confining part of the interaction does not play any role \cite{1105.4103}. On the other hand, excitations of quarks within a shell of width $\Lambda_{QCD}$ from the Fermi surface interact through infrared singular gluons at large $N_c$ and hence are confined.

In has been recently argued \cite{chiralspirals, chiralspirals-conf} that chiral symmetry can be broken in QyM through the formation of a translational non-invariant condensate that arises from the pairing between a quark with momentum p, and the hole formed by removing a quark with opposite momentum –p from the Fermi surface. The inhomogeneous condensate that forms in QyM is a linear combination of the chiral condensate $\langle\overline{\psi}\psi\rangle$, and a spin-one, isosinglet odd-parity condensate of $\langle\overline{\psi}\sigma^{0z}\psi\rangle$. Here z is the direction of motion of the wave. At each given patch of the Fermi surface, z is the direction perpendicular to the surface. The authors of Ref. \cite{chiralspirals} called this combination of two inhomogeneous condensates a Quarkyonic Chiral Spiral (QyCS). The $\langle\overline{\psi}\sigma^{0z}\psi\rangle$ component corresponds to the condensation of an electric dipole moment. The QyCS is then characterized by a spatial oscillation between chiral and electric dipole condensates that breaks parity and gives rise to an inhomogeneous electric field.

On the other hand, a common feature of heavy-ion collisions is the generation of strong magnetic fields that are produced in peripheral collisions by the positively charged ions moving at almost the speed of light. For the Au-Au collisions at the Relativistic Heavy Ion Collider (RHIC) at BNL the field produced is estimated to be $\sim 10^{19}G$ \cite{Fuku-review}. Even though this magnetic field decays quickly, it only decays to a tenth of the original value for a time scale of order of the inverse of the saturation scale at RHIC \cite{Fuku-review}, hence it may influence the properties of the QCD phases probed by the experiment. Strong magnetic fields will likely be also generated in future experiments planned at the Facility for Antiproton and Ion Research (FAIR) at GSI, the Nuclotron-Ion Collider Facility (NICA) at JINR, and the Japan Proton Accelerator Research Complex (JPARK) at JAERI, all of which intend to complement the experiments at RHIC by reaching regions of even higher densities and intermediate to low temperatures in the QCD phase map.

In this paper we discuss the effects of an external magnetic field on QyM. As will be seen below, the presence of a magnetic field of strength comparable to the square of the QCD scale gives rise to the formation of a second chiral spiral, given by a spatial oscillation between a pion condensate $\langle \overline{\psi}\gamma_{5} \psi\rangle$ and a spin-one condensate $\langle\overline{\psi}\gamma^{1}\gamma^{2}\psi\rangle$. The spin-one condensate corresponds to an inhomogeneous magnetic moment in the direction of the field.

\section{Quark Self-Energy in a Magnetic Field}

We are interested in studying the large $N_c$ properties of dense quark matter in the presence of a magnetic field B within a region where screening effects are not strong enough to eliminate confining and hence the quarkyonic phase can exist. Such a region can be defined by the condition $m_{D}\ll \Lambda_{QCD}\ll\mu$, with $m_{D}$ the screening mass and $\mu$ the quark chemical potential \cite{chiralspirals-conf}.

Our goal is to investigate the self-energy for quarks near the Fermi surface and in the presence of an external magnetic field. With that aim, we need to consider the Schwinger-Dyson equation for the quark self-energy $\Sigma_{\textrm{B}}(x,x')$ in a magnetic field,
\bea
   \Sigma_{\textrm{B}}(x,x')=g^2\gamma^\mu t^AG_{\textrm{B}}(x,x')\gamma^\nu t^BD^{AB}_{\mu\nu}(x-x')
\label{SD-xspace}
\eea
with $G_{\textrm{B}}(x,x')$ the full quark propagator in the presence of B and $D^{AB}_{\mu\nu}(x-x')$ the gluon propagator. Without loss of generality we choose the gauge $A_\mu=(0,0,-Bx_1,0)$ for the external  electromagnetic field. It produces a constant magnetic field in the $x_3$-direction.

Using Ritus's method \cite{ritus}, the transformation to momentum space of the quark propagator and self-energy can be done with the help of the matrix eigenfunctions
\bea
  \mathbb{E}_p&=&\sum_{\sigma=\pm1}E_{p\sigma}(x)\Delta(\sigma),
\label{epmatrices}
\eea
of the asymptotic states of the charged fermions in a uniform magnetic field, where $E_{p\sigma}(x)= \frac{(4\pi
|e_{f}B|)^{1/4}}{\sqrt{n!}}e^{i(p_0x^0+p_2x^2+p_3x^3)}D_n(\rho)$ are the corresponding eigenfunctions, with $D_n(\rho)$ the parabolic cylinder functions of argument $\rho=\sqrt{2|e_{f}B|}(x_1-p_2/e_{f}B)$, and $\Delta(\sigma)=\frac{I+i\sigma \gamma^1\gamma^2}{2}$ being the spin projectors. The $E_{p\sigma}(x)$ depend on the spin projection $\sigma$ and the Landau level $l\geq0$  through the non-negative integer $n=n(l,\sigma)\equiv l+\textrm{sgn}(e_{f}B)\frac{\sigma}{2}-\frac{1}{2}$. $e_f$ is the electric charge of flavor $f$.

In momentum space \Eq{SD-xspace} becomes
\bea
(2\pi)^4\hat{\delta}^4(p-p')\Sigma_{l}(\overline{p})\Pi_l
 &=&g^2t^At^B\int\frac{d^4q}{(2\pi)^4}
          D^{AB}_{\mu\nu}(q)
\nonumber \\
      &\times&   \sumint\frac{d^4p''}{(2\pi)^4}
         \int d^4x e^{-iq\cdot x}
         \overline{\mathbb{E}}_p(x)
         \gamma^\mu
         \mathbb{E}_{p''}(x)
\Pi_{l''}\tilde{G}^{l''}(\overline{p}'',\mu)
\nonumber\\
         &\times&
         \int d^4x'e^{iq\cdot x'}
         \overline{\mathbb{E}}_{p''}(x')
         \gamma^\nu
         \mathbb{E}_{p'}(x'),
\label{auto5}
\eea
where
\bea
  \Sigma(p,p')&=&
        \int d^4xd^4x'\overline{\mathbb{E}}_p(x)\Sigma(x,x')\mathbb{E}_p(x')
\nonumber \\
    &=&(2\pi)^4\hat{\delta}^{(4)}(p-p')\Pi_l\Sigma^{l}(\overline{p}),
 \label{ad10}
\eea
and we used
\bea
   D^{AB}_{\mu\nu}(x-x')=\int\frac{d^4q}{(2\pi^4)}
       e^{-iq\cdot (x-x')}D^{AB}_{\mu\nu}(q),
\label{auto3}
\eea
and
\bea
    G_{\textrm{B}}(x,x')=\sumint\frac{d^4p''}{(2\pi)^4}
           \mathbb{E}_{p''}(x)\Pi_{l''}
          G^{l''}(\overline{p}'',\mu)
           \overline{\mathbb{E}}_{p''}(x'),
\label{auto4}
\eea
Here $[G^{l}(\overline{p})]^{-1}=[\gamma_{\varrho}\overline{p}^{\varrho}+\Sigma^{l}(|\overline{p}|)]^{-1}$ and $\Sigma^{l}(\overline{p})=\Sigma_{\|}(|\overline{p}|)\gamma_{\|}\overline{p}^{\|}+\Sigma_{\perp}(|\overline{p}|)\gamma_{\perp}\overline{p}^{\perp}+\Sigma_m(|\overline{p}|)$, with $\overline{p}_{\mu}\equiv(p_0,0,\textrm{sgn}(e_fB)\sqrt{2|e_fB|l},p_3)$. The symbol  $\parallel$ indicates longitudinal components (0,3), while $\perp$ indicates the components perpendicular to the external magnetic field. $G^{l}(\overline{p},\mu)$ is defined by $G^{l}(\overline{p})$ with the replacement $p_0\rightarrow p_0+\mu$. In the above equations we used the following notation
${\sum\!\!\!\!\!\!\int} d^4p\equiv \sum_{l}\int dp_{0}dp_{2}dp_{3}$, $\overline{\mathbb{E}}_p(x)\equiv\gamma^0\mathbb{E}_p^\dagger\gamma^0$, and $ \hat{\delta}^{(4)}(p-p')\equiv\delta^{ll'}\delta(p_0-p'_0)\delta(p_2-p'_2)\delta(p_3-p'_3)$.
The projector $\Pi_l=\delta^{0l}\Delta(\textrm{sgn}(e_{f}B))+I(1-\delta^{0l})$ appears due to the lack of spin degeneracy of the lowest Landau level (LLL) \cite{leungwang}.

To consider confined excitations near the Fermi surface, we can use the Gribov-Zwanziger gluon propagator \cite{Zwanziger},
\bea
  D_{\mu\nu}^{AB}(q)= -i\delta_{AB}\frac{8\pi}{C_F} \frac{\tau}{(\vec{q}^{\,2})^2}g_{0\mu}g_{0\nu},
\label{G-Z propagt}
\eea
where $C_F=(N^{2}_{c}-1)/2N_{c}$ and the string tension $\tau\sim\Lambda^{2}_{QCD}$.

Since the strength of the magnetic fields produced in the heavy-ion collisions is comparable to the QCD scale, it makes sense to assume that the external magnetic field satisfies $e_fB\lesssim \mathcal{O}(\Lambda^{2}_{QCD})$. Summing in the color indexes and following derivations similar to those performed in \cite{leeleungng}, we can integrate in x, x' and p", and then transform to polar variables $\hat{q}_\perp=\sqrt{q^{2}_{1}+q^{2}_{2}}$, $\varphi \equiv \arctan(q_{2}/q_{1})$, integrate in $\varphi $, and finally perform the sums in Landau levels and spin indexes to arrive at
\bea
    &&\hspace{-0.2cm}\Sigma^{l}(\overline{p})\Pi_l=
     \frac{-ig^28\pi\tau}{ 2}\Pi_l\int \frac{d^2q_{\parallel} dq^2_{\perp}}{(2\pi)^3}
       e^{-\hat{q}^2_\perp}
    \frac{
    \gamma_0
    G^{l}(\overline{p-q},\mu)
    \gamma_0}
    {(\vec{q}^{\,2})^2}
\nonumber \\
\label{SD-3}
\eea
where $\overline{p-q}\equiv(p_0-q_0,0,\textrm{sgn}(e_fB)\sqrt{2|e_fB|l},p_3-q_3)$ and $\hat{q}^2_\perp=\frac{q^{2}_{1}+q^{2}_{2}}{2|e_fB|}$.

Because of the Landau quantization of the transverse momentum, the Fermi surface now is formed by a discrete set of circles, obtained from the intersection of the cylinders of radius $\sqrt{2|e_fB|l}$ with the spherical Fermi surface of the system at zero field, as represented in Fig.1a.

\begin{figure}[ht!]
\begin{center}
\label{fermi:surface}
\includegraphics[width=4cm]{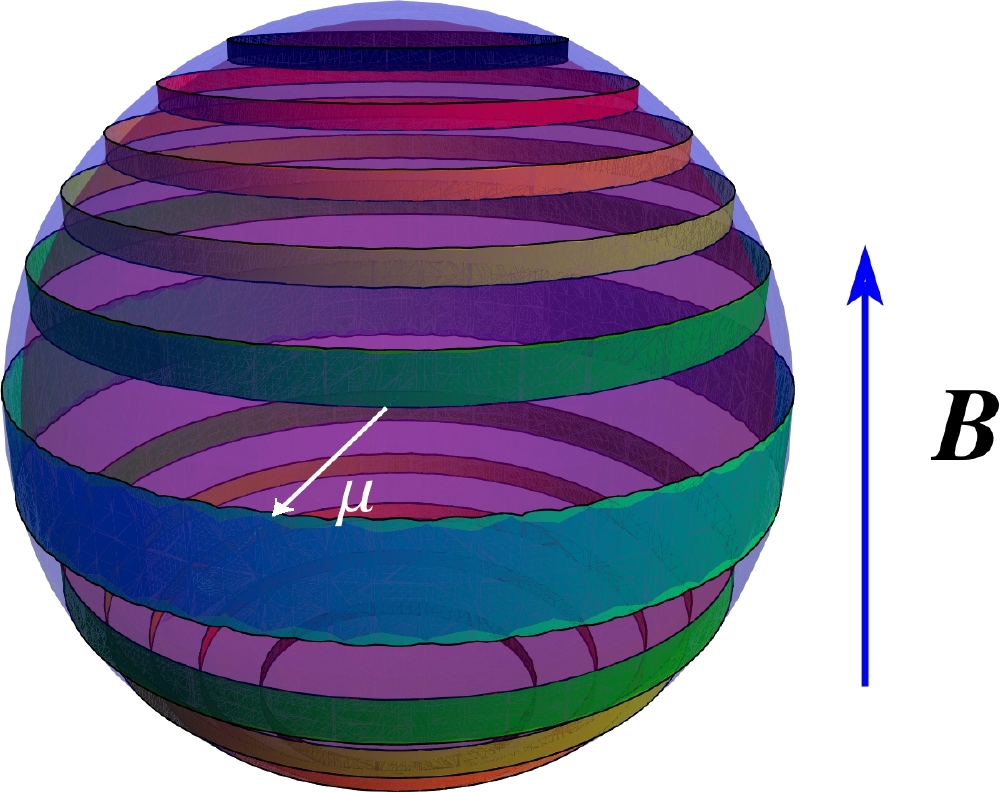}
\hspace{0.5in}
\label{fermi:patch}
\includegraphics[width=4cm]{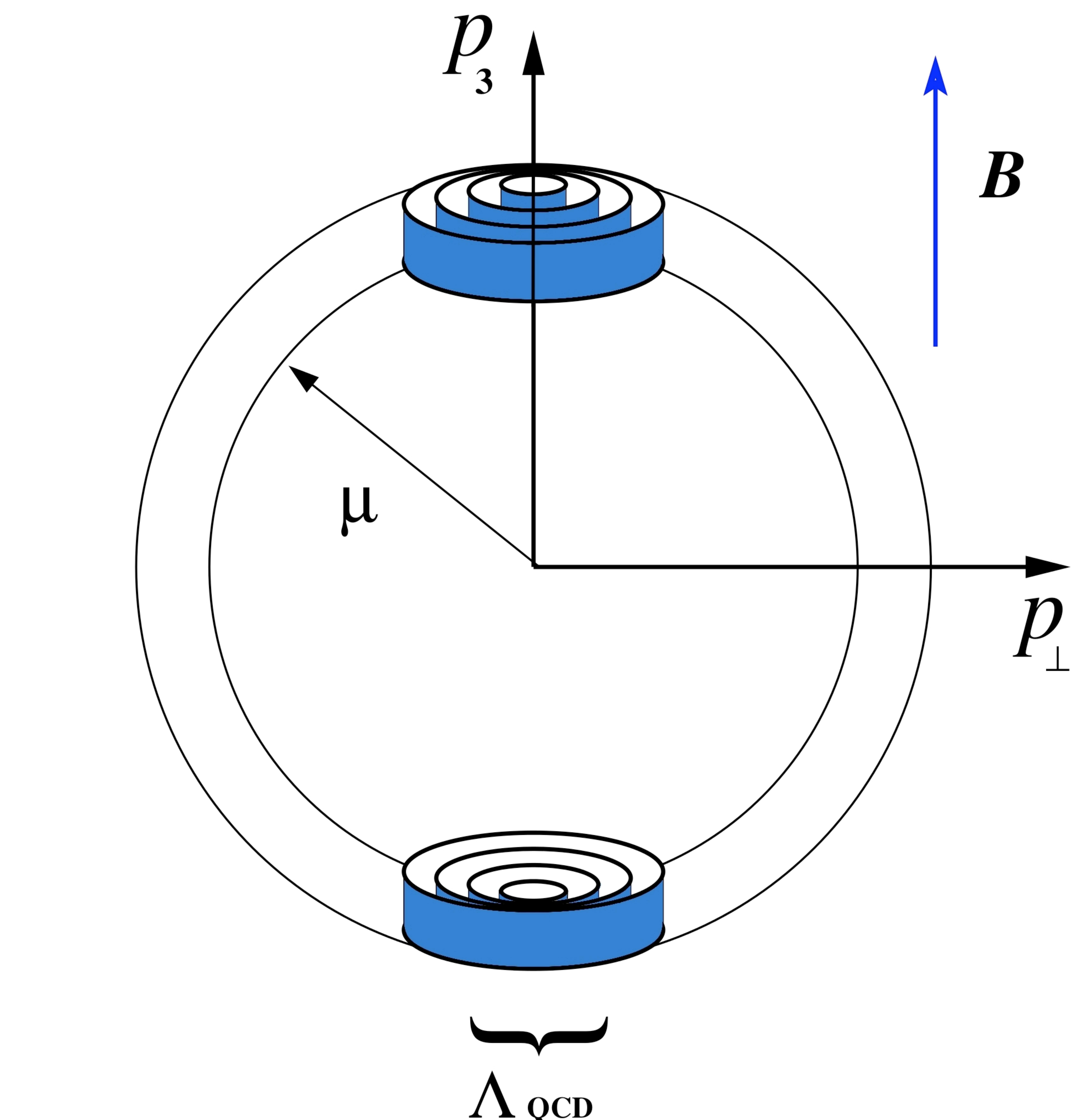}
\end{center}
\caption{a) Fermi surface in the presence of a magnetic field; 
b) Patches}
\label{fermi:figs}
\end{figure}

After transforming to Euclidean space, we can readily integrate in $q_{\perp}$, taking into account that the fermion propagator in \Eq{SD-3} does not depend on this variable,

 \bea \label{SD-4}
  \hspace{-0.2cm}\Sigma^l(\overline{p})\Pi_l
   &\simeq&
  \frac{N_cg_{2D}^2}{2}\Pi_l \int \frac{d^2q_{||}}{(2\pi)^2}
    \gamma_4
   G^{l}(\overline{p-q},\mu)
    \gamma_4
    \frac{1}{q_3^2}
\eea
Here we introduced a two-dimensional coupling constant $g^{2}_{2D}=\frac{4g^2\tau}{N_c}$. Since $p_4$ only enters in the rhs of (\ref{SD-4}) in the quark propagator, it can be eliminated with a change of variable. This implies that the self energy is actually independent of $p_4$.

\section{Magnetic Quarkyonic Chiral Spiral}

We are interesting in exploring the contribution to the self-energy from excitations about the Fermi surface on the patches of size $\sim\Lambda_{QCD}\ll\mu$, situated on opposite sites of the Fermi surface, and which lie perpendicular to the magnetic field direction (north and south pole patches indicated in Fig.1b). Quark excitations in these patches satisfy $p_3\approx \mu+\delta p_3$, with $\delta p_3\leq\Lambda_{QCD}\ll\mu$, for the momentum component parallel to the field, and $ p_{\perp}=\sqrt{2|e_{f}B|l}\leq\Lambda_{QCD}$ for the transverse component, so the number of Landau levels that can contribute to the patch is limited to $L=l_{\textit{max}}=[\frac{\Lambda^2_{QCD}}{2|e_{f}B|}]$, with [...] meaning integer part. These conditions, together with the strong infrared behavior in $q_3$ of the gluon propagator, allow us to neglect $\sqrt{2|e_{f}B|l}$ in $|\overline{p-q}|$ in the integrand and similarly drop the term $\Sigma_{\perp}\gamma_{2}\sqrt{2|e_{f}B|l}$ in $\Sigma^l$ in both sides of the equation.

As a consequence, any explicit dependence on the Landau level disappears in \Eq{SD-4}, except for the projector $\Pi_l$. If $l\neq0$, the two spin projections contribute and $\Pi_l$ becomes the identity matrix. In this case, \Eq{SD-4} looks exactly like the self-energy equation for two-dimensional QCD in the axial gauge \cite{chiralspirals,chiralspirals-conf}. \Eq{SD-4} is identical to the one found in QyM at zero magnetic field \cite{chiralspirals,chiralspirals-conf}, but since here we have one equation for each $l$ contained in the patch, the Landau level becomes a flavor index. Hence, there will be $L$ identical equations of the form
\bea \label{SD-l}
  \hspace{-0.2cm}\Sigma^l(p_3)
   \simeq
  \frac{N_cg_{2D}^2}{2}\int \frac{d^2q_{||}}{(2\pi)^2}
    \gamma_4
   G^{l}(iq_4+\mu,q_3-p_3)
    \gamma_4
    \frac{1}{q_3^2} \quad l=1,2,...L
\eea
 
 On the other hand, if $l=0$, $\Pi_l$ selects one of the two spin projections only, "up" for positively charged quarks ($e_f>0$), "down" for negatively charged quarks ($e_f<0$).

Then the original 4-dimensional theory for a quark field with electric charge $e_f$ in the presence of a magnetic field maps into a (1+1)-dimensional theory with flavor symmetry SU(2L)$\times$ U(1) described by the Lagrangian
\bea
   \emph{L}^{2D}_{eff}&=&
     \overline{\Phi}_0
     [i\Gamma^\mu(\partial_\mu+i g_{2D}A_\mu +\Gamma^0 \mu]\Phi_0
     \nonumber \\
    &+&\sum_{l=1}^{L}
     \overline{\Phi}_l[i\Gamma^\mu(\partial_\mu+ ig_{2D}A_\mu)+\Gamma^0\mu]\Phi_l -\frac{1}{2}trG^{2}_{\mu\nu}
\label{2DlagwithB}
\eea
Here the spinor fields are defined by $\Phi^{T}_0=( \varphi^{(0)}_{\uparrow},0)$ and $\Phi^{T}_l=( \varphi^{(l)}_{\uparrow},\varphi^{(l)}_{\downarrow})$, with $\uparrow$ and $\downarrow$ two flavors in the (1+1)-D theory that correspond to the up and down spin components of the spinor in the 4D theory. The 2D Dirac $\Gamma$ matrices are defined in term of the Pauli matrices as $\Gamma^0=\sigma^1$; $\Gamma^z=-i\sigma^2$; $\Gamma^5=\sigma^3$.

We can now perform the transformation of the quark fields
\bea
\Phi_{l}=exp(-i\mu z \Gamma_{5})\Phi'_{l} \quad\quad l=0,...L
\label{fieldtransf}
\eea
to eliminate the chemical potential, which actually remains in the theory through the anomaly of the baryon charge, and obtain
\bea
   \emph{L}^{2D}_{eff}&=&\sum_{l=0}^{L}
     \overline{\Phi}_l[i\Gamma^\mu(\partial_\mu+ ig_{2D}A_\mu)+\Gamma^0\mu]\Phi_l -\frac{1}{2}trG^{2}_{\mu\nu}
     \nonumber \\
     &=&\sum_{l=0}^{L}
     \overline{\Phi}'_{l}[i\Gamma^\mu(\partial_\mu+ ig_{2D}A_\mu)]\Phi'_{l} -\frac{1}{2}trG^{2}_{\mu\nu}
\label{transf2DlagwithB}
\eea

Following arguments along the lines of those discussed in Ref.\cite{chiralspirals}, we can argued that in the above theory chiral symmetry is broken through the formation of  a chiral condensate for each flavor $l$, giving rise to a total chiral condensate of the form $\langle \overline{\Phi}'\Phi' \rangle=\sum_{l=0}^{L} \langle \overline{\Phi}'_{l}\Phi'_{l} \rangle$. Performing the inverse transformation of (\ref{fieldtransf}), we obtain two inhomogeneous condensates, oscillating in z, and with equal amplitude forming a spiral in the chiral space: a quarkyonic chiral spiral, very similar to the one found at zero field
\bea
 \langle \overline{\Phi}\Phi \rangle
 =\cos(2\mu z)\langle \overline{\Phi}'\Phi' \rangle \quad \langle \overline{\Phi}\Gamma_5 \Phi \rangle
 =-i\sin(2\mu z)\langle \overline{\Phi}'\Phi' \rangle
\label{spirals1}
\eea

 One can easily check that the amplitude of the chiral spiral found in \cite{chiralspirals} can be expressed as a sum of condensates of spin-up and spin-down flavors, $\langle \overline{\Phi}'\Phi' \rangle=\langle \overline{\varphi}'_{\uparrow}\varphi'_{\uparrow} \rangle+\langle \overline{\varphi}'_{\downarrow}\varphi'_{\downarrow} \rangle $. It is now clear why a condensate of the form $\langle \overline{\Phi}'\tau_3 \Phi' \rangle=\langle \overline{\varphi}'_{\uparrow}\varphi'_{\uparrow} \rangle-\langle \overline{\varphi}'_{\downarrow}\varphi'_{\downarrow} \rangle $ cannot be present in the QyM at zero magnetic field (see Ref. \cite{chiralspirals} for the definition of the flavor matrix $\tau_3$ in the present context). There is nothing in the theory at zero magnetic field that can distinguish between a condensate of spin-up fields and one of spin-down fields, so physically these two condensates should be the same and they must cancel out in $\langle \overline{\Phi}'\tau_3 \Phi' \rangle$, in agreement with the results of Ref. \cite{chiralspirals}.

However, a magnetic field can change the value of $\langle \overline{\Phi}'\tau_3 \Phi' \rangle$ because the LLL flavor in the 2D theory has only one spin-flavor component. In consequence, two independent chiral condensates can be formed with the primed fields. In addition to a chiral condensate of the form
\bea
 \langle \overline{\Phi}'\Phi' \rangle 
 = \langle \overline{\varphi}^{'(0)}_{\uparrow}\varphi^{'(0)}_{\uparrow} \rangle +\sum_{l=1}^{L}[ \langle \overline{\varphi}^{'(l)}_{\uparrow}\varphi^{'(l)}_{\uparrow} \rangle+\langle \overline{\varphi}^{'(l)}_{\downarrow}\varphi^{'(l)}_{\downarrow} \rangle]
\label{Ch-cond-decomp1}
\eea
we also have now
\bea
 \langle \overline{\Phi}'\tau_3 \Phi' \rangle
 = \langle \overline{\varphi}^{'(0)}_{\uparrow}\varphi^{'(0)}_{\uparrow} \rangle +\sum_{l=1}^{L}[ \langle \overline{\varphi}^{'(l)}_{\uparrow}\varphi^{'(l)}_{\uparrow} \rangle-\langle \overline{\varphi}^{'(l)}_{\downarrow}\varphi^{'(l)}_{\downarrow} \rangle]
\label{Ch-cond-decomp2}
\eea
which is always different from zero due to the LLL contribution, even if the two spin-flavor terms in the sum cancel out. The new condensate gives rise to the following second chiral spiral
\bea
 \langle \overline{\Phi}\tau_3\Phi \rangle
 =\cos(2\mu z)\langle \overline{\Phi}'\tau_3\Phi' \rangle \quad \langle \overline{\Phi}\tau_3\Gamma_5 \Phi \rangle
 =-i\sin(2\mu z)\langle \overline{\Phi}'\tau_3 \Phi' \rangle
\label{spirals2}
\eea

Going back to the quark fields in the (3+1)-dimensional theory, the two chiral spirals in the presence of a magnetic field are
\bea
 \langle \overline{\psi}\psi \rangle
 =\Delta_1\cos(2\mu z) \quad \quad \langle \overline{\psi}\gamma^0\gamma^3\psi \rangle
 =\Delta_1\sin(2\mu z)
\label{4Dspirals1}
\eea
\bea
 \langle \overline{\psi}\gamma^1\gamma^2\psi \rangle
 =\Delta_2 \cos(2\mu z) \quad \quad \langle \overline{\psi}\gamma^5\psi \rangle
 =\Delta_2 \sin(2\mu z)
\label{4Dspirals2}
\eea

\section{Conclusions}

In this paper we have considered QyM in a magnetic field of strength close to, but not larger than the QCD scale. We showed that in leading order in the large $N_c$ limit, a (3+1)-dimensional QCD theory with one flavor can be mapped into a (1+1)-dimensional QCD theory with $2L+1$ flavors and global symmetry SU(2L)$\times$ U(1). 

Due to the contribution of the LLL, two independent chiral spirals are formed in the Fermi surface's patches perpendicular to the magnetic field. One, similar to the spiral condensate that forms at zero field, is a combination of a chiral condensate and an electric dipole condensate, rotating in the chiral space and varying along the direction parallel to the field. The other chiral spiral is a combination of a condensate of magnetic moment and a pion condensate, also varying in the direction parallel to the field. Both parity and time-reversal symmetries are broken in the system. 

The spontaneous generation of inhomogeneous condensates with electric and magnetic dipole moments may lead to interesting observational implications. A study of those potential consequences in dense environments like the cores of neutron stars or the planed high-density heavy-ion collisions experiments will be addressed in future works.

\textbf{Acknowledgments:} The work of VI and EJF has been supported
in part by DOE Nuclear Theory Grant No. DE-SC0002179. The authors thank Larry McLerran and Toru Kojo for encouragement and enlightening discussions.

\end{document}